\documentstyle[amssymb,rmp,aps]{revtex}

\begin{document}
\title{Relaxation of the Phase In an Inhomogeneous Bose Gas}
\author{V. S. Babichenko}
\address{R.S.C. ''Kurchatov Institute'' Kurchatov sq. 46, Moscow 123182, Russia.}
\date{\today}
\maketitle

\begin{abstract}
An effective action is obtained of a Bose gas in the bulk separated into two
regions by a strong external potential depending on the single coordinate.
The main attention is focused on the relaxation of the difference between
phases of the weakly coupling condensates of the different bulk domains
separated from each other by the external potential.
\end{abstract}

The experimental realization of the Bose-Einstein condensation at ultralow
temperatures in atomic vapors [1] provides an example of the systems in
which the approximation of the weakly non-ideal gas is well applicable. This
fact is connected with the small density of particles in the systems
concerned. The realization of such systems gives the possibility of the
experimental investigations of the macroscopic quantum phenomenons of
different types. Recently, the manifestations of the macroscopic quantum
phase and its behavior are studied actively [1-8], [11].

The investigation of the kinetic phenomena due to the relaxation of the
order parameter is of a doubtless interest for the study of macroscopic
quantum phenomena. In the present work we study the spatial relaxation of
the phase of the order parameter in the inhomogeneous Bose gas with the weak
coupling between different spatial regions due to a barrier tunneling.

The system concerned in the present work is a Bose gas of the small density
in the bulk separated into two regions by a strong external potential $%
U\left( \overrightarrow{r}\right) $ which depends on the x coordinate and
does not depend on $\overrightarrow{r}_{\perp }=\left( y,z\right) $, i.e., $%
U\left( \overrightarrow{r}\right) =U\left( x\right) $. For simplicity, we
assume that the external potential has a rectangular shape $U\left( x\right)
=U_{0}$ for $-d<x<d$ and $U\left( x\right) =0$ beyond this region of the x
coordinate. The height of the external potential $U_{0}$ is supposed to be
the largest energy parameter in the system, in particular, $U_{0}>>\mu $
where $\mu $ is the chemical potential $\mu =n\lambda $, $\lambda $ is the
scattering amplitude of Bose particles, and n is the density of the Bose
gas. Due to this assumption the interaction between particles of the Bose
gas in the region of the influence of the external potential, i.e., in the
region $-d<x<d$, can be neglected.

The left and the right domains of the bulk are supposed to have the same
temperature $T$ smaller than the Bose-condensation temperature $T_{c}$. For
simplicity, we suppose that in the initial state the density of the left and
the right domains of the bulk are the same, but the phases of the left and
the right Bose condensates are different. This assumption means that at the
initial time moment there is nonzero current from one side of the bulk to
another. This non-equilibrium initial state will relax to the equilibrium
one which has the same densities and the same phases of the condensates for
both sides of the bulk. This relaxation process is studied in the present
work.

Note that in the case of superconductors the similar initial non-equilibrium
state results in the Josephson oscillations with a small damping [9], [10].
The essential distinction of the Bose gas consisting of neutral atoms from
superconductors in which the Cooper pairs represent the charged objects is
the presence of the gapless excitation spectrum. The presence of low energy
excitations results in the change of the character of the initial state
relaxation making the relaxation essentially faster.

In the case of the homogeneous in y-z plane difference between the phases of
the condensates the excitations radiated during the relaxation process have
the one-dimensional character. The one-dimensional character of the radiated
excitations results in a divergency of the small momentum correlators of
these excitations. Hence, the consideration of the relaxation process to
second order in the tunneling amplitude is not correct in contrast to the
case of superconductors [10] and requires more accurate analysis.

\section{Perturbation theory in the tunneling amplitude}

As the introduction, we consider the perturbation theory in the amplitude of
the tunneling under the external potential barrier. The Hamiltonian of the
Bose gas with the tunneling term is

\begin{eqnarray}
\widehat{H} &=&\int\limits_{U\left( x\right) =0}d^{3}r\left\{ \widehat{\psi }%
^{+}\left( \overrightarrow{r}\right) \left( -\frac{1}{2m}\overrightarrow{%
\nabla }^{2}-\mu \right) \widehat{\psi }\left( \overrightarrow{r}\right) +%
\frac{\lambda }{2}\left( \widehat{\psi }^{+}\left( \overrightarrow{r}\right) 
\widehat{\psi }\left( \overrightarrow{r}\right) \right) ^{2}\right\} + 
\eqnum{1.1} \\
&&+\int d^{3}r_{1}d^{3}r_{2}\left\{ \widehat{\psi }^{+}\left( 
\overrightarrow{r}_{1}\right) \widehat{T}_{\overrightarrow{r}_{1}%
\overrightarrow{r}_{2}}\widehat{\psi }\left( \overrightarrow{r}_{2}\right) +%
\widehat{\psi }^{+}\left( \overrightarrow{r}_{2}\right) \widehat{T}_{%
\overrightarrow{r}_{1}\overrightarrow{r}_{2}}^{\ast }\widehat{\psi }\left( 
\overrightarrow{r}_{1}\right) \right\}  \nonumber
\end{eqnarray}
The form of the kernel of the integral operator $T_{\overrightarrow{r}_{1}%
\overrightarrow{r}_{2}}$ in the case of the external potential independent
of $\overrightarrow{r}_{\perp }$ can be taken as

\begin{equation}
T_{\overrightarrow{r}_{1}\overrightarrow{r}_{2}}=\gamma _{x_{1},x_{2}}\delta
\left( \overrightarrow{r}_{\perp 1}-\overrightarrow{r}_{\perp 2}\right) 
\text{ \ \ \ \ \ \ where \ \ }x_{1}>0\text{ \ \ and \ \ }x_{2}<0  \eqnum{1.2}
\end{equation}
The kernel $\gamma _{x_{1},x_{2}}$ is localized near \ $x_{1}=x_{2}=0$
within the coordinate x of about $a$ where $a\lesssim \xi $, $\xi $ being
the correlation length of the weakly interacting Bose gas $\xi $ $=\left(
n\lambda \right) ^{-1/2}$. In the case of the slow variation of the operator 
$\widehat{\psi }\left( \overrightarrow{r}\right) $ compared with the scale $%
\xi $ the kernel $\gamma _{x_{1},x_{2}}$ can be taken in the form

\begin{equation}
\gamma _{x_{1},x_{2}}=\gamma _{0}\delta \left( x_{1}\right) \delta \left(
x_{2}\right)  \eqnum{1.3}
\end{equation}
The constant $\gamma _{0}$ is the small quantity due to the large magnitude
of the external potential and obeys the inequality $\gamma _{0}<<\mu \xi $.
The tunneling amplitude (1.3) will be obtained below and has the form $%
\gamma _{0}=\frac{1}{2m}n\xi \left( \varkappa \sinh \left( 2\varkappa
d\right) \right) ^{-1}$ where $\varkappa =\sqrt{2mU_{0}}$. Form (1.3) of the
tunneling amplitude differs from that used in the work [11]. This difference
is due to the behavior of the Fourier components of the tunneling amplitude $%
\gamma _{k_{1x},k_{2x}}$ for the small momenta $k_{1x},k_{2x}$. In contrast
to the present work, in the work [11] the tunneling amplitude $\gamma
_{k_{1x},k_{2x}}$ vanishes $\gamma _{k_{1x},k_{2x}}\rightarrow 0$ provided
that the momenta $k_{1x},k_{2x}$ tends to zero $k_{1x},k_{2x}\rightarrow 0$,
this amplitude being nonzero for the condensate-condensate tunneling. This
distinction has the far-reaching consequences, namely, nonzero value of $%
\gamma _{k_{1x},k_{2x}}$ for $k_{1x},k_{2x}\rightarrow 0$ results in\ the
divergency of lower orders of the perturbation theory in the tunneling
amplitude. This problem is discussed in detail below.

The characteristic frequency of the phase difference evolution between the
right and left condensates is proportional to the small tunneling amplitude $%
\gamma _{0}$. The characteristic momenta of particles corresponding to this
frequency are small compared with the scale of the order of $1/\xi $. This
means that the slow relaxation process of the phase difference is described
by the slow variation in the spatial and temporal Fourier components of
operators $\widehat{\psi }_{\alpha }\left( \overrightarrow{r},t\right) $.
Index $\alpha $ is equal to $\alpha =1,2$ where 1,2\ denotes the right-hand
(x%
\mbox{$>$}%
0) and the left-hand (x%
\mbox{$<$}%
0) sides of the bulk, respectively. Therefore, the model form of the kernel $%
\gamma _{x_{1},x_{2}}$ given by Eq. (1.3) corresponds to the purpose of the
work, namely, to describe the relaxation of the phase difference.

Due to the existence of the condensates in the left-hand and the right-hand
sides of the bulk the operators $\widehat{\psi }_{\alpha }\left( 
\overrightarrow{r},t\right) $ can be represented in the well-known form

\begin{equation}
\widehat{\psi }_{\alpha }\left( \overrightarrow{r},t\right) =\Psi _{\alpha
}\left( t\right) +\widehat{\chi }_{\alpha }\left( \overrightarrow{r},t\right)
\eqnum{1.4}
\end{equation}

\begin{equation}
\Psi _{\alpha }\left( t\right) =\sqrt{n_{0}}e^{i\varphi _{\alpha }\left(
t\right) }  \eqnum{1.5}
\end{equation}
The fields $\Psi _{\alpha }\left( t\right) $\ are the condensate fields for
the right-hand ($\alpha =1$) and the left-hand ($\alpha =2$) sides of the
bulk. The value $n_{0}$ is the Bose condensate density and $\varphi _{\alpha
}\left( t\right) $ are the phases of the right-hand ($\alpha =1$) and the
left-hand ($\alpha =2$) Bose-condensates. The operators $\widehat{\chi }%
_{\alpha }\left( \overrightarrow{r},t\right) $ are the annihilation
operators of non-condensate particles. As has been written above, the main
problem is the investigation of the time evolution of the magnitude $\Delta
\varphi \left( t\right) =$ $\varphi _{1}\left( t\right) -\varphi _{2}\left(
t\right) $. For this purpose, we will obtain the effective action for the
phases $\varphi _{1}\left( t\right) $ and $\varphi _{2}\left( t\right) $.

At first, we consider the perturbation theory in the tunneling amplitude for
the small phase difference $\Delta \varphi \left( t\right) $. The diagrams
for the renormalized tunnelling amplitude of the condensate particles have
the form of the vertex with two external lines (as for the self-energy part)
corresponding to the condensate field. The quantities corresponding to the
incoming and outgoing dotted lines are $\Psi _{\alpha }$ and $\overline{\Psi 
}_{\alpha }$, respectively. To analyse the relaxation process, we use the
Keldysh diagram technique with the Schwinger-Keldysh two-time contour [14],
[15].

The reason of the impossibility to consider the lower orders of the
perturbation theory in the tunneling amplitude is the divergency of the
lower order diagrams when the external frequency $\omega $ tends to zero $%
\omega \rightarrow 0$. To demonstrate this, we consider the frequency
Fourier transformation of one of second-order tunneling vertex

\begin{equation}
\Gamma _{11}^{\left( 2\right) }=\gamma _{0}g\left( \omega \right) \gamma _{0}
\eqnum{1.6}
\end{equation}
where the function $g\left( \omega \right) $ denotes the Green function of
the non-condensate fields [13] with the momentum $\overrightarrow{k}_{\perp
}=0$ and the same x-coordinates

\begin{equation}
g\left( \omega \right) =G_{1}\left( \omega ,\overrightarrow{k}_{\perp
}=0;x_{1}=x_{2}=d\right) =G_{2}\left( \omega ,\overrightarrow{k}_{\perp
}=0;x_{1}=x_{2}=-d\right)  \eqnum{1.7}
\end{equation}
The Green functions $G_{\alpha }\left( t-t^{\prime },\overrightarrow{r}-%
\overrightarrow{r}^{\prime }\right) $ $\left( \alpha =1,2\right) $ are the
Belyaev Green functions of the non-condensate particles of the homogeneous
interacting Bose gas. The momentum $\overrightarrow{k}_{\perp }$
corresponding to the Green function $g\left( \omega \right) $ vanishes $%
\overrightarrow{k}_{\perp }=0$ due to the homogeneity of the condensate
field in the y and z-directions. The Green functions $G_{\alpha }\left(
\omega ,\overrightarrow{k}\right) $ and, correspondingly, $g\left( \omega
\right) $ in the Keldysh diagrammatic technique has three components in
triangle representation, namely, retarded $G^{\left( R\right) }$, advanced $%
G^{\left( A\right) }$\ and kinetic$\ G^{\left( K\right) }$\ Green functions
[14], [15]. In the case of small frequencies $\omega <<\mu $\ the retarded
Green function $g^{\left( R\right) }\left( \omega \right) $, for example,
can be written in the form

\begin{equation}
g^{\left( R\right) }\left( \omega \right) =\int \frac{dp_{x}}{2\pi }\left( 
\frac{\mu }{\left( \omega +i\delta \right) ^{2}-c^{2}p_{x}^{2}}\right)  
\eqnum{1.8}
\end{equation}
The small momentum asymptotic of G in Eq. (1.8) is used due to the smallness
of the external frequency $\omega $ $\left( \omega <<\mu \right) $. The main
contribution into integral (1.8) is provided by the momenta $p_{x}\thicksim
\omega /c<<\mu /c$. Note that integral (1.8) has a singularity and,
therefore, the magnitude of this integral is determined by the bypass around
the  pole. The calculation of the integral (1.8) gives

\begin{equation}
g^{\left( R\right) }\left( \omega \right) =-i\frac{\mu }{c}\frac{1}{2\omega }
\eqnum{1.9}
\end{equation}
This behavior of the function $g^{\left( R\right) }\left( \omega \right) $
means that for the sufficiently small magnitudes of $\omega $, the
consideration of some lower orders of the perturbation theory is not correct
and the analysis of diagrams in all orders is necessary.

The diagrams of odd order of this perturbation theory results in the
renormalization of the bare tunneling amplitude $\gamma _{0}$. The sum of
the odd order diagrams having one incoming and one outgoing line is denoted
by the vertex $\Gamma _{\alpha \beta }$ $\left( \alpha \neq \beta \right) $
where, as above, the indices $\alpha ,\beta =1,2$ denote the right-hand or
the left-hand sides of the bulk. The sum of the odd order diagrams having
two incoming external lines or two outgoing external lines is denoted by the
vertex $\widehat{\Gamma }_{\alpha \beta }$ $\left( \alpha \neq \beta \right) 
$ according to the notations of the self-energy part in the Belyaev diagram
technique for the Bose systems [12], [13]. The term of the effective action
corresponding to this renormalized tunneling amplitude has the form

\begin{eqnarray*}
S_{tunn} &=&\oint \oint dtdt^{\prime }\overline{\Psi }_{\alpha }\left(
t\right) \Gamma _{\alpha \beta }\left( t,t^{\prime }\right) \Psi _{\beta
}\left( t^{\prime }\right) + \\
&&+\oint \oint dtdt^{\prime }\left\{ \Psi _{\alpha }\left( t\right) \widehat{%
\Gamma }_{\alpha \beta }\left( t,t^{\prime }\right) \Psi _{\beta }\left(
t^{\prime }\right) +\overline{\Psi }_{\alpha }\left( t\right) \widehat{%
\Gamma }_{\alpha \beta }\left( t,t^{\prime }\right) \overline{\Psi }_{\beta
}\left( t^{\prime }\right) \right\}
\end{eqnarray*}
The time contour in this expression is the Schwinger-Keldysh two-time
contour. Note, that in the even orders of the tunneling amplitude the
renormalized vertex $\Gamma _{\alpha \beta }$ has the coincident external
indexes $\alpha =\beta $. These vertexes give the contribution to the
self-energy part of the particles in one of the two subvolumes, in
particular, the imaginary parts of these vertexes describe the relaxation
process.

\ The renormalized vertexes connected with the dressed tunneling amplitude
can be expressed by means of the Green functions taking into account the
tunneling

\begin{equation}
\Gamma _{11}=\gamma _{0}G_{22}\gamma _{0}  \eqnum{1.10}
\end{equation}

\[
\widehat{\Gamma }_{11}=\gamma _{0}F_{22}\gamma _{0} 
\]

\[
\Gamma _{12}=\gamma _{0}+\gamma _{0}G_{21}\gamma _{0} 
\]

\[
\widehat{\Gamma }_{12}=\gamma _{0}F_{21}\gamma _{0} 
\]
The system of the Dyson equations\ for $G_{\alpha \beta }$ and $F_{\alpha
\beta }$ can be written as

\begin{equation}
G_{22}=G_{2}+G_{2}\gamma _{0}G_{12}+F_{2}\gamma _{0}F_{12}  \eqnum{1.11}
\end{equation}

\[
F_{22}=F_{2}+F_{2}\gamma _{0}G_{12}+\widetilde{G}_{2}\gamma _{0}F_{12} 
\]

\[
G_{12}=G_{1}\gamma _{0}G_{22}+F_{1}\gamma _{0}F_{22} 
\]

\[
F_{12}=\widetilde{G}_{1}\gamma _{0}F_{22}+F_{1}\gamma _{0}G_{22} 
\]
Here $G_{\alpha }$ and $F_{\alpha }$ are the Belyaev Green functions of
non-condensate particles of the homogeneous interacting Bose gas. The Green
function $G_{\alpha }$ has one incoming and one outgoing end and the Green
function $F_{\alpha }$ has two outgoing ( or two incoming) ends. The Green
function $\widetilde{G}_{\alpha }$ denotes the Green function $G_{\alpha }$
with the opposite frequency $\omega $ and momentum $\overrightarrow{p}$,
i.e., $\widetilde{G}_{\alpha }\left( \omega ,\overrightarrow{p}\right)
=G_{\alpha }\left( -\omega ,-\overrightarrow{p}\right) $. This system of
equations is written in the assumption of the small magnitude of the
difference between phases of the left-hand and right-hand condensates. In
the case of the small momenta, which is the most interesting for us, the
Green functions $G_{\alpha }$ and $F_{\alpha }$ obeys the equality $%
G_{\alpha }=-F_{\alpha }$. In the case of the small momenta the solutions of
Eqs. (1.11) can be simplified and the tunneling vertexes take the form

\begin{equation}
\Gamma _{11}=\frac{g\left( \omega \right) \gamma _{0}^{2}}{1-4\gamma
_{0}^{2}g^{2}\left( \omega \right) }  \eqnum{1.12}
\end{equation}

\[
\widehat{\Gamma }_{11}=-\frac{g\left( \omega \right) \gamma _{0}^{2}}{%
1-4\gamma _{0}^{2}g^{2}\left( \omega \right) } 
\]

\[
\Gamma _{12}=\gamma _{0}+\frac{2g^{2}\left( \omega \right) \gamma _{0}^{3}}{%
1-4\gamma _{0}^{2}g^{2}\left( \omega \right) } 
\]

\[
\widehat{\Gamma }_{12}=\frac{-2g^{2}\left( \omega \right) \gamma _{0}^{3}}{%
1-4\gamma _{0}^{2}g^{2}\left( \omega \right) } 
\]
In the limit $\omega \rightarrow 0$ these vertexes can be written as

\begin{eqnarray}
\Gamma _{11}^{\left( R\right) } &\rightarrow &-\frac{1}{4g\left( \omega
\right) }=-i\frac{c}{2\mu }\omega  \eqnum{1.13} \\
\widehat{\Gamma }_{11}^{\left( R\right) } &\rightarrow &\frac{1}{4g\left(
\omega \right) }=i\frac{c}{2\mu }\omega  \nonumber
\end{eqnarray}

\begin{eqnarray*}
\Gamma _{12}^{\left( R\right) } &\rightarrow &\frac{1}{2}\gamma _{0} \\
\widehat{\Gamma }_{12}^{\left( R\right) } &\rightarrow &\frac{1}{2}\gamma
_{0}
\end{eqnarray*}

Note that the bare tunneling amplitude $\gamma _{0}$ is eliminated from the
renormalized vertexes $\Gamma _{11}$ , $\widehat{\Gamma }_{11}$, and just
these vertexes describe the relaxation of the phase (difference). This fact
results in the large value of the relaxation rate.

The more rigorous and detailed consideration of the problem in the case of
an arbitrary phase difference for the infinitely large size of the system in
the x-direction is represented below. The assumption about the size of the
system is accepted to focus the attention on the consideration of the
relaxation process.

\section{The tunneling action}

Below the approach to the problem of the phase difference relaxation without
using the perturbation theory in the tunneling amplitude is developed. For
this purpose the effective action for the boundary values of the Bose field
phases is obtained by the integration over the volume components of these
fields with the fixed values of them at the left-hand and the right-hand
sides of the external potential boundary. Integrating over these fields we
take into account the inhomogeneous behavior of the condensate fields in the
left-hand and the right-hand sides of the volume. The necessity of the
consideration of the inhomogeneous condensate behavior even for the small
momentums of the excitations is conditioned by the fact that the self
consistent inhomogeneous condensate additional field makes the situation in
some sense similar to the case of the scattering on the external potential
having the discrete energy level which is very close to zero. This can
results, for example, in the transmission coefficient of the order of unit
including the case of the very small compared with $\mu $ energies [16]. The
obtained effective action contains the relaxation part which is proportional
to the first power of the frequency of the phase. For the small frequencies
the relaxation part is found to be much larger than the usual kinetic part
of the phase dynamics which is proportional to the second power of the
frequency. The consideration is developed in the framework of the
Schwinger-Keldysh technique with the time reversing contour [14], [15].

The generating functional of the system can be written in the form

\begin{equation}
Z=\int D\psi D\overline{\psi }e^{i\left( S+S_{j}\right) }  \eqnum{2.1}
\end{equation}
The action S of the inhomogeneous non-equilibrium Bose gas is given by

\begin{equation}
S=\oint dt\int d^{3}r\left\{ \overline{\psi }\left[ i\partial _{t}+\mu
-U\left( x\right) \right] \psi -\frac{1}{2}\left( \overrightarrow{\nabla }%
\overline{\psi }\right) \left( \overrightarrow{\nabla }\psi \right) -\frac{%
\lambda }{2}\left( \overline{\psi }\psi \right) ^{2}\right\}  \eqnum{2.2}
\end{equation}
and term $S_{j}$ reads as

\begin{equation}
S_{j}=\oint dt\int d^{3}r\left( \overline{\psi }j+\overline{j}\psi \right) 
\eqnum{2.3}
\end{equation}
Here $j$ and $\overline{j}$ are the infinitely small sources. The Planck
constant $\hbar $ and mass m of a Bose particle are set equal to unity $%
\hbar =m=1$.

Due to the large magnitude of potential $U_{0}$ in the region x$\in \left[
-d,d\right] $ the interaction between Bose particles can be neglected and
the integral for Z over the fields $\psi ,\overline{\psi }$ takes the
Gaussian form in this region of the x-coordinate. In this connection the
integral for Z over the fields $\psi \left( x,\overrightarrow{r}_{\perp
};t\right) ,\overline{\psi }\left( x,\overrightarrow{r}_{\perp };t\right) $
for x$\in \left[ -d,d\right] $ with the fixed magnitudes of these fields at
the boundary of the region of the nonzero external potential can be
calculated. Moreover, the characteristic scale of the $\psi $-field
variation in the time and in the $\overrightarrow{r}_{\perp }$ space is
supposed to be much larger than $1/U_{0}$ and $1/\sqrt{U_{0}}$,
respectively. It is these fields that describe the slow relaxation process
possessing the characteristic frequency proportional to the small tunneling
coefficient. Due to this assumption the tunneling amplitude of the $\psi $
fields has the local character for the time variable t and space variables $%
\overrightarrow{r}_{\perp }$.

The magnitudes of the $\psi $-fields at the boundary of the region of the
nonzero external potential are denoted as

\begin{eqnarray}
\psi _{1}\left( x=d,\overrightarrow{r}_{\perp };t\right) &=&\psi _{s1}\left( 
\overrightarrow{r}_{\perp };t\right)  \eqnum{2.4} \\
\psi _{2}\left( x=-d,\overrightarrow{r}_{\perp };t\right) &=&\psi
_{s2}\left( \overrightarrow{r}_{\perp };t\right)  \nonumber \\
\overline{\psi }_{1}\left( x=d,\overrightarrow{r}_{\perp };t\right) &=&%
\overline{\psi }_{s1}\left( \overrightarrow{r}_{\perp };t\right)  \nonumber
\\
\overline{\psi }_{2}\left( x=-d,\overrightarrow{r}_{\perp };t\right) &=&%
\overline{\psi }_{s2}\left( \overrightarrow{r}_{\perp };t\right)  \nonumber
\end{eqnarray}
The calculation of the integral for Z over the $\psi $-fields in the region x%
$\in \left[ -d,d\right] $ gives

\begin{equation}
Z=\int D\psi D\overline{\psi }e^{i\left( S_{vol}+S_{surf}+S_{j}\right) } 
\eqnum{2.5}
\end{equation}
The part of the action $S_{vol}$ is

\begin{equation}
S_{vol}=\oint dt\int\limits_{U\left( x\right) =0}dx\int d^{2}r_{\perp
}\left\{ \overline{\psi }\left( i\partial _{t}+\mu \right) \psi -\frac{1}{2}%
\left( \overrightarrow{\nabla }\overline{\psi }\right) \left( 
\overrightarrow{\nabla }\psi \right) -\frac{\lambda }{2}\left( \overline{%
\psi }\psi \right) ^{2}\right\}  \eqnum{2.6}
\end{equation}
where the region of the integration over the x-coordinate is the sum of the
regions -L%
\mbox{$<$}%
x%
\mbox{$<$}%
-d and d%
\mbox{$<$}%
x%
\mbox{$<$}%
L.

The part of the action $S_{surf}$ can be represented as a sum of two terms $%
S_{surf}^{\left( \rho \right) }$ and $S_{surf}^{\left( tunn\right) }$ [17]

\begin{equation}
S_{surf}=S_{surf}^{\left( \rho \right) }+S_{surf}^{\left( tunn\right) } 
\eqnum{2.7}
\end{equation}

\begin{equation}
S_{surf}^{\left( \rho \right) }=-\frac{\varkappa }{2}\coth \left( 2\varkappa
d\right) \oint dt\int d^{2}r_{\perp }\left\{ \mid \psi _{s1}\left( 
\overrightarrow{r}_{\perp };t\right) \mid ^{2}+\mid \psi _{s2}\left( 
\overrightarrow{r}_{\perp };t\right) \mid ^{2}\right\}  \eqnum{2.8}
\end{equation}

\begin{equation}
S_{surf}^{\left( tunn\right) }=\frac{\varkappa }{2\sinh \left( 2\varkappa
d\right) }\oint dt\int d^{2}r_{\perp }\left\{ \overline{\psi }_{s1}\left( 
\overrightarrow{r}_{\perp };t\right) \psi _{s2}\left( \overrightarrow{r}%
_{\perp };t\right) +\overline{\psi }_{s2}\left( \overrightarrow{r}_{\perp
};t\right) \psi _{s1}\left( \overrightarrow{r}_{\perp };t\right) \right\} 
\eqnum{2.9}
\end{equation}
where the magnitude $\varkappa $ is $\varkappa =\sqrt{2U_{0}}$. Later on we
assume that the width d\ of the potential barrier is sufficiency large so
that the inequality $\varkappa d>>1$ takes place and, thus, $\sinh \left(
2\varkappa d\right) \thickapprox \cosh \left( 2\varkappa d\right)
\thickapprox \frac{1}{2}e^{2\varkappa d}>>1$. The term $S_{surf}^{\left(
tunn\right) }$ describes the tunneling between the right-hand and left-hand
sides of the bulk. From the form of this tunneling part of the action it can
easily be seen that the tunneling amplitude for the condensate particles and
for the low energy non-condensate particles has the same nonzero magnitude
in contrast with [11].

Our goal is the calculation of the functional integral for Z over the $\psi $%
-fields in the bulk, i.e., in the region x$\in \left( -L,-d\right) \cup
\left( d,L\right) $ with the fixed values at the boundary of the region of
the nonzero external potential, and, thus, obtaining the effective action
for the fields at the boundary $\psi _{s1}$, $\psi _{s2}$.

\section{The saddle-point approximation}

Later on, it is convenient to change the variables in the following way

\begin{eqnarray}
\overrightarrow{r} &\rightarrow &\xi \overrightarrow{r}  \eqnum{3.1} \\
t &\rightarrow &\frac{1}{\mu }t  \nonumber
\end{eqnarray}
where $\xi =1/\sqrt{\mu }$. Thus, the system of units which we use later
measures the quantities of the dimension of length in units of $\xi $ and
the quantities of the dimension of energy in units of $\mu $.

Moreover, it is convenient to represent the fields $\psi $ in the
modulus-phase form

\begin{equation}
\psi =\rho e^{i\varphi }\text{; \ \ \ \ \ \ }\overline{\psi }=\rho
e^{-i\varphi }  \eqnum{3.2}
\end{equation}
The integral over the fields $\psi $ in the right-hand and the left-hand
sides of the bulk can be calculated within the framework of the saddle-point
approximation. The saddle-point magnitude $\psi _{cl}\left( x,%
\overrightarrow{r}_{\perp };t\right) $ of the field $\psi $ obeys the
Gross-Pitaevskii equation with the boundary conditions

\begin{eqnarray}
\psi _{cl}\left( x=d,\overrightarrow{r}_{\perp };t\right) &=&\psi
_{s1}\left( \overrightarrow{r}_{\perp };t\right)  \eqnum{3.3} \\
\psi _{cl}\left( x=-d,\overrightarrow{r}_{\perp };t\right) &=&\psi
_{s2}\left( \overrightarrow{r}_{\perp };t\right)  \nonumber
\end{eqnarray}

In the present consideration we use the saddle-point approximation with
respect to the modulus of the field $\psi $, i.e., $\rho $. The fluctuations
of the modulus field $\rho $ in the vicinity of the saddle-point magnitude $%
\rho ^{\left( cl\right) }$\ are taken into account within the framework of
the perturbation theory. At the same time, the phase $\varphi $ is
considered exactly. The reason of such consideration is connected with the
one-dimensional character of the excitations which are radiated during the
relaxation of the difference between the phases of the left-hand and
right-hand condensates. The saddle-point magnitudes $\rho ^{\left( cl\right)
}$ of the fields $\rho $ have the form

\begin{eqnarray}
\rho _{1}^{\left( cl\right) }\left( x,\overrightarrow{r}_{\perp };t\right)
&=&\sqrt{n_{0}}F_{1}\left( x,\overrightarrow{r}_{\perp };t\right) \text{ \ \
\ \ \ \ \ \ \ \ \ \ \ \ }\left( x>d\right)  \eqnum{3.4} \\
\rho _{2}^{\left( cl\right) }\left( x,\overrightarrow{r}_{\perp };t\right)
&=&\sqrt{n_{0}}F_{2}\left( x,\overrightarrow{r}_{\perp };t\right) \text{ \ \
\ \ \ \ \ \ \ \ \ \ \ \ }\left( x<-d\right)  \nonumber
\end{eqnarray}
where $n_{0}$ is the condensate density in the bulk far from the region of
the nonzero external potential and

\begin{eqnarray}
F_{1}\left( x,\overrightarrow{r}_{\perp };t\right) &=&\tanh \left(
x-d+X_{1}\left( \overrightarrow{r}_{\perp };t\right) \right)  \eqnum{3.5} \\
F_{2}\left( x,\overrightarrow{r}_{\perp };t\right) &=&\tanh \left(
x+d+X_{2}\left( \overrightarrow{r}_{\perp };t\right) \right)  \nonumber
\end{eqnarray}
The fields $X_{1,2}\left( \overrightarrow{r}_{\perp };t\right) $ are
determined by the boundary magnitudes of the field $\rho ^{\left( cl\right)
}\left( x=\pm d\right) $

\begin{eqnarray}
\rho _{s1}\left( \overrightarrow{r}_{\perp };t\right) &=&\rho _{1}^{\left(
cl\right) }\left( x=d,\overrightarrow{r}_{\perp };t\right) =\sqrt{n_{0}}%
\tanh \left( X_{1}\left( \overrightarrow{r}_{\perp };t\right) \right) 
\eqnum{3.6} \\
\rho _{s2}\left( \overrightarrow{r}_{\perp };t\right) &=&\rho _{2}^{\left(
cl\right) }\left( x=-d,\overrightarrow{r}_{\perp };t\right) =\sqrt{n_{0}}%
\tanh \left( X_{2}\left( \overrightarrow{r}_{\perp };t\right) \right) 
\nonumber
\end{eqnarray}

Due to the large value of the external potential U$_{0}$ (or $\varkappa >>1$%
) the essential contribution into the integral for Z is given by the fields
with the small values of the modulus at the boundary $\mid \psi
_{s1,2}\left( \overrightarrow{r}_{\perp };t\right) \mid <<\sqrt{n_{0}}$.
This means that the fields $X_{1,2}$ can be considered as the fields with
the small values $\mid X_{1,2}\mid <<1$ and, thus, these fields can be
expressed in the simple form in terms of the boundary values of the fields $%
\rho _{s1,2}$ $\left( \overrightarrow{r}_{\perp };t\right) $

\[
X_{1,2}\left( \overrightarrow{r}_{\perp };t\right) =\frac{1}{\sqrt{n_{0}}}%
\rho _{s1,2}\left( \overrightarrow{r}_{\perp };t\right) 
\]

\section{Density fluctuations}

For the consideration of the fluctuations of the modulus, the field $\rho $
is represented as

\begin{equation}
\rho =\rho ^{\left( cl\right) }+\sqrt{n_{0}}\delta \rho  \eqnum{4.1}
\end{equation}
The boundary conditions are taken into account in the definition of the
field $\rho ^{\left( cl\right) }$ and, therefore, the fluctuations of the
modulus $\sqrt{n_{0}}\delta \rho $ should obey the zero boundary conditions

\begin{equation}
\delta \rho _{1}\left( x=d,\overrightarrow{r}_{\perp };t\right) =\delta \rho
_{2}\left( x=-d,\overrightarrow{r}_{\perp };t\right) =0  \eqnum{4.2}
\end{equation}
Within the framework of the saddle-point approximation the inequality $\mid
\delta \rho \mid <<\rho ^{\left( cl\right) }/\sqrt{n_{0}}=F$ should hold
for. Substituting (4.1) into the action (2.6) and taking into account the
inequality $\mid \delta \rho \mid <<\rho ^{\left( cl\right) }/\sqrt{n_{0}}=F$%
, we obtain \ 

\begin{equation}
S_{vol}=S_{vol}^{\left( 0\right) }+\delta S_{vol}  \eqnum{4.3}
\end{equation}
where

\begin{equation}
S_{vol}^{\left( 0\right) }=n_{0}\xi ^{3}\oint dt\int\limits_{U\left(
x\right) =0}dx\int d^{2}r_{\perp }\left\{ -F^{2}\left( \partial _{t}\varphi
\right) +F^{2}-\frac{1}{2}F^{4}-\frac{1}{2}\left( \overrightarrow{\nabla }%
F\right) ^{2}\right\}  \eqnum{4.4}
\end{equation}

\begin{equation}
\delta S_{vol}=n_{0}\xi ^{3}\oint dt\int\limits_{U\left( x\right) =0}dx\int
d^{2}r_{\perp }\left\{ 
\begin{array}{c}
-2F\delta \rho \left( \partial _{t}\varphi \right) -\frac{1}{2}F^{2}\left( 
\overrightarrow{\nabla }\varphi \right) ^{2}+\delta \rho \widehat{H}%
_{1}\delta \rho + \\ 
+\frac{1}{2}\delta \rho \overrightarrow{\nabla }_{\perp }^{2}\delta \rho
\end{array}
\right\}  \eqnum{4.5}
\end{equation}
where the function $F$ is obtained by Eq. (3.5) and the operator $\widehat{H}%
_{1}$ reads

\begin{equation}
\widehat{H}_{1}=-\frac{1}{2}\nabla _{x}^{2}-1+3F^{2}  \eqnum{4.6}
\end{equation}

The spectrum of the operator $\widehat{H}_{1}$ can be found [19] and has the
form

\begin{equation}
\lambda _{0}=0\text{, \ \ }\lambda _{1}=\frac{3}{2}\text{\ , \ }\lambda
_{k}=2+\frac{1}{2}k^{2}\text{\ }  \eqnum{4.7}
\end{equation}
Note that the zero mode self-energy $\lambda _{0}$ of operator $\widehat{H}%
_{1}$ should be eliminated due to zero boundary conditions (4.2). At the
same time the second discrete self-energy $\lambda _{1}$ and the
self-energies of the continuous spectrum $\lambda _{k}$ are not changed
essentially by boundary conditions (4.2). In Eqs. (4.4), (4.5) the
fluctuations of modulus $\delta \rho $ are taken into account within an
accuracy to second order \ due to inequality $\delta \rho <<F$\ and the
phase is taken into account exactly. The possibility of the consideration of
the fluctuations $\delta \rho $ within an accuracy to second order results
from the gap character of the spectrum of these fluctuations, in contrast to
the spectrum of the fluctuations of the phase which is gapless.

The contribution of the densify fluctuations into the generation functional
Z is determined by action $\delta S_{vol}$ and has the form of the Gaussian
integral over $\delta \rho $. The integration over $\delta \rho $ in Z can
be performed and we obtain

\begin{equation}
Z=\int DXD\varphi e^{i\left( S_{vol}^{\left( 0\right) }+S^{\left( \varphi
\right) }+S_{surf}+S_{j}\right) }  \eqnum{4.8}
\end{equation}

\begin{equation}
S^{\left( \varphi \right) }=n_{0}\xi ^{3}\oint dt\int d^{3}\overrightarrow{r}%
d^{3}\overrightarrow{r}^{\prime }\left\{ 
\begin{array}{c}
\left( F\left( \partial _{t}\varphi \right) \right) _{\overrightarrow{r}%
}\left( \widehat{H}_{1}-\frac{1}{2}\overrightarrow{\nabla }_{\perp
}^{2}\right) _{\overrightarrow{r},\overrightarrow{r}^{\prime }}^{-1}\left(
F\left( \partial _{t}\varphi \right) \right) _{\overrightarrow{r}^{\prime }}-
\\ 
-\frac{1}{2}F^{2}\left( \overrightarrow{\nabla }\varphi \right) ^{2}\delta 
\left[ \overrightarrow{r}-\overrightarrow{r}^{\prime }\right]
\end{array}
\right\}  \eqnum{4.9}
\end{equation}

The evolution of the difference between the phases of the condensates in the
right-hand and the left-hand sides of the bulk is a slow process. The
inverse characteristic time of this process is proportional to the tunneling
amplitude $\gamma =\frac{1}{2}n_{0}\left( \varkappa \sinh \left( 2\varkappa
d\right) \right) ^{-1}$ and has the small value. Therefore, the momenta $%
\overrightarrow{k}_{\perp }$ of the fluctuations corresponding to these
fluctuation frequencies should have small value $k_{\perp }\xi <<1$ (or in
our system of units where $\xi =1$, $k_{\perp }<<1$). As a result, due to
the gap character of the self-energies of operator $\widehat{H}_{1}$ in the
case of zero boundary conditions for $\delta \rho $ in the direction $%
\overrightarrow{r}_{\perp }$ we can neglect the term $\frac{1}{2}%
\overrightarrow{\nabla }_{\perp }^{2}$ compared with $\widehat{H}_{1}$ in
the first summand in the action (4.9).

The operator $\widehat{H}_{1}$ can be represented in the factorized form

\begin{equation}
\widehat{H}_{1}=\frac{1}{2}\widehat{L}_{1}^{+}\widehat{L}_{1}  \eqnum{4.10}
\end{equation}
where the operators $\widehat{L}_{1}$, $\widehat{L}_{1}^{+}$ have the form

\begin{eqnarray}
\widehat{L}_{1} &=&\nabla _{x}+2F  \eqnum{4.11} \\
\widehat{L}_{1}^{+} &=&-\nabla _{x}+2F  \nonumber
\end{eqnarray}
Due to the possibility to represent $\widehat{H}_{1}$ in the factorized form
and to neglect $\overrightarrow{\nabla }_{\perp }^{2}$ in the first summand
of Eq. (4.9), the action $S^{\left( \varphi \right) }$ can be written as

\begin{equation}
S^{\left( \varphi \right) }=n_{0}\xi ^{3}\oint dt\int\limits_{U\left(
x\right) =0}dx\int d^{2}\overrightarrow{r}_{\perp }\left\{ 2\left[ \left( 
\widehat{L}_{1}^{+}\right) ^{-1}\left( F\partial _{t}\varphi \right) \right]
^{2}-\frac{1}{2}F^{2}\left( \overrightarrow{\nabla }\varphi \right)
^{2}\right\}  \eqnum{4.12}
\end{equation}
After the integration by parts over x in the second term of the action $%
S^{\left( \varphi \right) }$\ (4.12) this action can be transformed to the
sum of two terms

\begin{equation}
S^{\left( \varphi \right) }=S_{vol}^{\left( \varphi \right) }+\widetilde{S}%
_{surf}^{\left( \varphi \right) }  \eqnum{4.13}
\end{equation}
where

\begin{equation}
S_{vol}^{\left( \varphi \right) }=n_{0}\xi ^{3}\oint dt\int\limits_{U\left(
x\right) =0}dx\int d^{2}\overrightarrow{r}_{\perp }\left\{ 
\begin{array}{c}
2\left[ \left( \widehat{L}_{1}^{+}\right) ^{-1}\left( F\partial _{t}\varphi
\right) \right] ^{2}-\left( F\varphi \right) \widehat{H}_{2}\left( F\varphi
\right) - \\ 
-\frac{1}{2}F^{2}\left( \overrightarrow{\nabla }_{\perp }\varphi \right) ^{2}
\end{array}
\right\}  \eqnum{4.14}
\end{equation}

\begin{equation}
\widetilde{S}_{surf}^{\left( \varphi \right) }=-\frac{1}{2}n_{0}\xi
^{3}\oint dt\int d^{2}\overrightarrow{r}_{\perp }\left\{ \left[ F^{2}\varphi
_{1}\left( \nabla _{x}\varphi _{1}\right) \right] \mid _{x=d}^{x=L}+\left[
F^{2}\varphi _{2}\left( \nabla _{x}\varphi _{2}\right) \right] \mid
_{x=-L}^{x=-d}\right\}  \eqnum{4.15}
\end{equation}

\begin{equation}
\widehat{H}_{2}=-\frac{1}{2}\nabla _{x}^{2}-1+F^{2}  \eqnum{4.16}
\end{equation}
The term $\widetilde{S}_{surf}^{\left( \varphi \right) }$\ of the action $%
S^{\left( \varphi \right) }$\ is a surface one. The region of the
integration in this surface term is the surface of the right-hand and the
left-hand sides of the bulk.

Note that the operator $\widehat{H}_{2}$ can be represented in the
factorized form as well as operator $\widehat{H}_{1}$

\begin{equation}
\widehat{H}_{2}=\frac{1}{2}\widehat{L}_{2}^{+}\widehat{L}_{2}  \eqnum{4.17}
\end{equation}
where

\begin{eqnarray}
\widehat{L}_{2} &=&-\nabla _{x}+\frac{\left( \nabla _{x}F\right) }{F} 
\eqnum{4.18} \\
\widehat{L}_{2}^{+} &=&\nabla _{x}+\frac{\left( \nabla _{x}F\right) }{F} 
\nonumber
\end{eqnarray}
It can easily be obtained that the operators $\widehat{L}_{1}$ and $\widehat{%
L}_{2}$ obey the following equations

\begin{equation}
\widehat{L}_{1}\widehat{H}_{2}\widehat{L}_{1}^{+}=\frac{1}{2}\widehat{L}_{1}%
\widehat{L}_{2}^{+}\widehat{L}_{2}\widehat{L}_{1}^{+}=\frac{1}{2}\nabla
_{x}\left( \nabla _{x}^{2}-4\right) \nabla _{x}  \eqnum{4.19}
\end{equation}

\begin{equation}
\widehat{L}_{1}\widehat{L}_{1}^{+}=2\widehat{H}_{2}+4  \eqnum{4.20}
\end{equation}

\section{The surface terms of the action}

The part of the action $S_{vol}^{\left( 0\right) }$ which consists of the
modulus terms alone and does not have the phase fluctuations can be
transformed to the surface form. This can easily be seen by the substitution
of equation (3.5) for F into the action $S_{vol}^{\left( 0\right) }$ (4.4).
Taking into account the smallness of X (X%
\mbox{$<$}%
\mbox{$<$}%
1), we obtain

\begin{eqnarray}
&&n_{0}\xi ^{3}\oint dt\int\limits_{U\left( x\right) =0}dx\int d^{2}r_{\perp
}\left\{ F^{2}-\frac{1}{2}F^{4}-\frac{1}{2}\left( \nabla _{x}F\right) ^{2}-%
\frac{1}{2}\left( \overrightarrow{\nabla }_{\perp }F\right) ^{2}\right\} 
\eqnum{5.1} \\
&=&n_{0}\xi ^{3}\oint dt\int d^{2}r_{\perp }\left\{ X-\frac{1}{3}\left( 
\overrightarrow{\nabla }_{\perp }X\right) ^{2}\right\}  \nonumber
\end{eqnarray}
Note that taking into account the smallness of X and assuming the slowness
of $X$ compared with the scale $\xi $ in the $\overrightarrow{r}_{\perp }$
space (the reason of this assumption has been discussed above), we can
neglect the term $\left( \overrightarrow{\nabla }_{\perp }X\right) ^{2}$ .

The term of the action $S_{vol}^{\left( 0\right) }$ depending on the
fluctuation of the phase can be transformed to the surface form in the
assumption of the slowness of the field $\varphi $ compared with the scale $%
\xi $.

\begin{eqnarray}
&&n_{0}\xi ^{3}\oint dt\int\limits_{U\left( x\right) =0}dx\int d^{2}r_{\perp
}\left\{ F^{2}\left( \partial _{t}\varphi \right) \right\} =  \eqnum{5.2} \\
&=&n_{0}\xi ^{3}\oint dt\int\limits_{U\left( x\right) =0}dx\int
d^{2}r_{\perp }\left\{ \left( F^{2}-1\right) \left( \partial _{t}\varphi
\right) \right\} =  \nonumber \\
&=&n_{0}\xi ^{3}\oint dt\int d^{2}r_{\perp }\left\{ X\left( \partial
_{t}\varphi _{s}\right) \right\}  \nonumber
\end{eqnarray}

As a result, the sum of $S_{vol}^{\left( 0\right) }$ and $S_{surf}$ is given
by

\begin{equation}
S_{vol}^{\left( 0\right) }+S_{surf}=n_{0}\xi ^{3}\oint dt\int d^{2}r_{\perp
}\left\{ X-\frac{\varkappa }{2}X^{2}-X\left( \partial _{t}\varphi
_{s}\right) +\frac{\varkappa }{\sinh \left( 2\varkappa d\right) }X^{2}\cos
\left( \Delta \varphi _{s}\right) \right\}  \eqnum{5.3}
\end{equation}
where $\Delta \varphi _{s}=\varphi _{s1}-\varphi _{s2}$ is the difference
between the magnitudes of the phase fields at the boundary of the region of
the nonzero external potential.

Due to the large value of $n_{0}\xi ^{3}>>1$ the saddle-point approximation
for the integral over X in the generation functional can be used. As can
readily be seen, the saddle-point value of X is equal to X$_{0}$

\begin{equation}
X_{0}=\frac{1}{\varkappa }  \eqnum{5.4}
\end{equation}
After integrating Z over the fluctuations of X near the saddle-point value X$%
_{0}$ and taking into account (4), we obtain the surface part of the action
in the form

\begin{equation}
\widetilde{\widetilde{S}}_{surf}^{\left( \varphi \right) }=n_{0}\xi
^{3}\oint dt\int d^{2}r_{\perp }\left\{ -\frac{1}{2\varkappa }\left(
\partial _{t}\varphi _{s}\right) ^{2}+\frac{1}{\varkappa \sinh \left(
2\varkappa d\right) }\cos \left( \Delta \varphi _{s}\right) \right\} 
\eqnum{5.5}
\end{equation}
Combining two surface terms $\widetilde{S}_{surf}^{\left( \varphi \right) }$
(5.15) and $\widetilde{\widetilde{S}}_{surf}^{\left( \varphi \right) }$ (5),
we obtain the surface term in the form

\begin{equation}
S_{surf}^{\left( \varphi \right) }=\widetilde{S}_{surf}^{\left( \varphi
\right) }+\widetilde{\widetilde{S}}_{surf}^{\left( \varphi \right) } 
\eqnum{5.6}
\end{equation}

\section{The effective action for the phase}

Due to the smallness of the fluctuations of X we replace X with the saddle
point value X$_{0}$ in the field F and denote this field as F$^{\left(
0\right) }$ where

\begin{equation}
F^{\left( 0\right) }\left( x\right) =\tanh \left( \frac{\mid x\mp d\mid }{%
\xi }+X_{0}\right)  \eqnum{6.1}
\end{equation}
The function F$^{\left( 0\right) }$ is substituted in the action S$%
_{vol}^{\left( \varphi \right) }$ defined by Eq. (4.13) instead of F. The
same procedure refers also to the operators $\widehat{L}_{1}^{+}$, and $%
\widehat{L}_{1}$. After this replacement the bulk part of the action S$%
_{vol}^{\left( \varphi \right) }$ can be written as

\begin{equation}
S_{vol}^{\left( \varphi \right) }=n_{0}\xi ^{3}\oint dt\int\limits_{U\left(
x\right) =0}dx\int d^{2}r_{\perp }\left\{ 
\begin{array}{c}
2\left[ \left( \widehat{L}_{1}^{\left( 0\right) +}\right) ^{-1}\left(
F^{\left( 0\right) }\partial _{t}\varphi \right) \right] ^{2}- \\ 
-\left( F^{\left( 0\right) }\varphi \right) \widehat{H}_{2}^{\left( 0\right)
}\left( F^{\left( 0\right) }\varphi \right) -\frac{1}{2}\left( F^{\left(
0\right) }\right) ^{2}\left( \overrightarrow{\nabla }_{\perp }\varphi
\right) ^{2}
\end{array}
\right\}  \eqnum{6.2}
\end{equation}
where the operators $\widehat{L}_{1}^{\left( 0\right) +}$ and $\widehat{H}%
_{2}^{\left( 0\right) }$ have the form of $\widehat{L}_{1}^{+}$ and $%
\widehat{H}_{2}$ with the substitution $F\rightarrow F^{\left( 0\right) }$.
Since the action $S_{vol}^{\left( \varphi \right) }$ has such form, that it
is convenient to introduce the field $\eta $ instead of the field $\varphi $

\begin{equation}
\eta =\left( \widehat{L}_{1}^{\left( 0\right) +}\right) ^{-1}\left(
F^{\left( 0\right) }\varphi \right)  \eqnum{6.3}
\end{equation}
The transformation yields

\begin{equation}
S_{vol}^{\left( \varphi \right) }=n_{0}\xi ^{3}\oint dt\int\limits_{U\left(
x\right) =0}dx\int d^{2}r_{\perp }\left\{ 
\begin{array}{c}
2\left( \partial _{t}\eta \right) ^{2}-\eta \left( \widehat{L}_{1}^{\left(
0\right) }\widehat{H}_{2}^{\left( 0\right) }\widehat{L}_{1}^{\left( 0\right)
+}\right) \eta - \\ 
-\frac{1}{2}\left( \overrightarrow{\nabla }_{\perp }\eta \right) \left( 
\widehat{L}_{1}^{\left( 0\right) }\widehat{L}_{1}^{\left( 0\right) +}\right)
\left( \overrightarrow{\nabla }_{\perp }\eta \right)
\end{array}
\right\}  \eqnum{6.4}
\end{equation}
Using (4.19) and (4.20), the action $S_{vol}^{\left( \varphi \right) }$ can
be written as

\begin{equation}
S_{vol}^{\left( \varphi \right) }=2n_{0}\xi ^{3}\oint dt\int\limits_{U\left(
x\right) =0}dx\int d^{2}r_{\perp }\left\{ 
\begin{array}{c}
\left( \partial _{t}\eta \right) ^{2}-\frac{1}{4}\eta \left[ \nabla
_{x}\left( \nabla _{x}^{2}-4\right) \nabla _{x}\right] \eta - \\ 
-\left( \overrightarrow{\nabla }_{\perp }\eta \right) \left( \frac{1}{2}%
\widehat{H}_{2}^{\left( 0\right) }+1\right) \left( \overrightarrow{\nabla }%
_{\perp }\eta \right)
\end{array}
\right\}  \eqnum{6.5}
\end{equation}
In the case of the slow spatial variation of the field $\eta $ along the x
axis the action (6.5) can be transformed to

\begin{equation}
S_{vol}^{\left( \varphi \right) }=2n_{0}\xi ^{3}\oint dt\int\limits_{U\left(
x\right) =0}dx\int d^{2}r_{\perp }\left\{ \left( \partial _{t}\eta \right)
^{2}+\eta \left( \nabla _{x}^{2}\right) \eta -\frac{1}{2}\left( 
\overrightarrow{\nabla }_{\perp }\eta \right) \left( 1+\left( F^{\left(
0\right) }\right) ^{2}\right) \left( \overrightarrow{\nabla }_{\perp }\eta
\right) \right\}  \eqnum{6.6}
\end{equation}
The transition to the Fourier transformation in the $\overrightarrow{r}%
_{\perp }$ plane gives

\begin{equation}
S_{vol}^{\left( \varphi \right) }=2n_{0}\xi ^{3}\oint dt\int\limits_{U\left(
x\right) =0}dx\int \frac{d^{2}k_{\perp }}{\left( 2\pi \right) ^{2}}\left\{
\eta _{-\overrightarrow{k}_{\perp }}\left[ -\partial _{t}^{2}+\nabla
_{x}^{2}-\overrightarrow{k}_{\perp }^{2}-\frac{1}{2}\overrightarrow{k}%
_{\perp }^{2}\left( \nabla _{x}F^{\left( 0\right) }\right) \right] \eta _{%
\overrightarrow{k}_{\perp }}\right\}  \eqnum{6.7}
\end{equation}

The surface terms in the action $\widetilde{S}_{surf}^{\left( \varphi
\right) }$ (4.15) can easily be expressed in terms of the field $\eta $.
Using the connection between the fields $\varphi $ and $\eta $ (6.3) having
the form

\begin{equation}
\varphi =\frac{1}{F^{\left( 0\right) }}\left( -\nabla _{x}+2F^{\left(
0\right) }\right) \eta  \eqnum{6.8}
\end{equation}
we express $\nabla _{x}\varphi $ in terms of $\eta $

\begin{equation}
\nabla _{x}\varphi =\frac{1}{F^{\left( 0\right) }}\left( -\nabla _{x}+\frac{%
1+\left( F^{\left( 0\right) }\right) ^{2}}{F^{\left( 0\right) }}\right)
\nabla _{x}\eta  \eqnum{6.9}
\end{equation}

The characteristic frequency $\omega _{0}$ of the phase time evolutions is
proportional to the tunneling amplitude $\omega _{0}\thicksim \gamma $,
where the tunneling amplitude $\gamma =\left( \varkappa \sinh \left(
2\varkappa d\right) \right) ^{-1}$. The tunneling amplitude $\gamma $ is the
small value being much smaller than $\frac{1}{\varkappa }=F^{\left( 0\right)
}\left( \pm d\right) <<1$, i.e., $\mu $\ in the usual system of units. Thus,
the characteristic frequency of the field $\eta $ is small and, therefore, $%
\nabla _{x}\eta $ is small compared with $\eta $ due to the connection $%
\omega \thicksim k_{x}$ on the mass surface for the small $\omega $ (Eq.
6.7). In this case Eq. (6.9) goes over into

\begin{equation}
\nabla _{x}\varphi =\frac{1+\left( F^{\left( 0\right) }\right) ^{2}}{\left(
F^{\left( 0\right) }\right) ^{2}}\nabla _{x}\eta  \eqnum{6.10}
\end{equation}
From Eqs. (6.8) and (6.10) due to the inequality $\omega _{0}\thicksim
k_{x}^{\left( 0\right) }\thicksim \gamma <<F^{\left( 0\right) }\left( \pm
d\right) $ ($k_{x}^{\left( 0\right) }$ is the characteristic momentum
corresponding to $\omega _{0}$) we obtain

\begin{equation}
\varphi \left( d\right) =2\eta \left( d\right)  \eqnum{6.11}
\end{equation}

\[
\nabla _{x}\varphi \left( d\right) =\frac{1}{\left( F^{\left( 0\right)
}\left( d\right) \right) ^{2}}\nabla _{x}\eta \left( d\right) 
\]

\[
\varphi \left( L\right) =2\eta \left( L\right) 
\]

\[
\nabla _{x}\varphi \left( L\right) =2\nabla _{x}\eta \left( L\right) 
\]
From these equalities the surface terms in the Eq. (4.15) for $\widetilde{S}%
_{surf}^{\left( \varphi \right) }$ can be written in terms of the $\eta $
field in the form

\begin{equation}
\frac{1}{2}\left[ \left( F^{\left( 0\right) }\right) ^{2}\varphi \nabla
_{x}\varphi \right] \left( d\right) =\left[ \eta \nabla _{x}\eta \right]
\left( d\right)  \eqnum{6.12}
\end{equation}

\section{The integral over the phase}

Using expressions (6.5-6.7) for action $S_{vol}^{\left( \varphi \right) }$,
the integral over the field $\varphi \left( x,\overrightarrow{r}_{\perp
};t\right) $ with the fixed values $\varphi _{s}\left( \overrightarrow{r}%
_{\perp };t\right) $ on the surface of the region of the nonzero external
potential can be calculated. Note that the phases of the left-hand and the
right-hand condensates are homogeneous in the $\overrightarrow{r}_{\perp }$
plane and do not depend on $\overrightarrow{r}_{\perp }$ due to the
translational invariance of the system in these directions. Therefore, the
surface phase fields $\varphi _{s}\left( \overrightarrow{r}_{\perp
};t\right) $ depending on $\overrightarrow{r}_{\perp }$ correspond to
non-condensate particles and can be considered as perturbations. We analyse
the relaxation of a difference in phases in zero order of these
perturbations and do not consider them in this work, i.e., we put $\varphi
_{s1}\left( t\right) $ and $\varphi _{s2}\left( t\right) $\ depending on the
time variable alone.

The action $S_{vol}^{\left( \varphi \right) }$ (6.5) has the Gaussian form.
In this connection, in order to calculate the integral over the bulk
components of the field $\eta $ for the small $\omega $ and $\overrightarrow{%
k}$, we should find the solution of the equation

\begin{equation}
\left[ \omega ^{2}+\nabla _{x}^{2}-\overrightarrow{k}_{\perp }^{2}+\frac{1}{2%
}\overrightarrow{k}_{\perp }^{2}\left( \nabla _{x}F^{\left( 0\right)
}\right) \right] \eta =0  \eqnum{7.1}
\end{equation}
as a function of the boundary values of the field $\eta $ and put this
solution into the action $S_{vol}^{\left( \varphi \right) }$. We supposed
that the boundary values of the phase field $\varphi $, i.e., $\varphi _{s}$
and $\varphi _{L}$, are known. In the case of the boundary magnitudes of the
field $\varphi $ and, hence, of the field $\eta $ independent on $%
\overrightarrow{r}_{\perp }$ we look for the solution of this equation for $%
\overrightarrow{k}_{\perp }=0$, i.e.,

\begin{equation}
\left[ \omega ^{2}+\nabla _{x}^{2}\right] \eta =0  \eqnum{7.2}
\end{equation}

As was mentioned above, to describe the relaxation, we use the
Keldysh-Schwinger technique with the two-time contour [14], [15]. In this
technique the field $\eta $ has two components corresponding to two branches
of the time contour (quadrangular representation). It is convenient to
introduce the components of the field $\eta $ as a sum and a difference of
the two components corresponding to the upper and to the lower branches of
the time contour (triangular representation) [14], [15]. In Eqs. (7.1),
(7.2) the frequency $\omega $ is a complex quantity $\omega \rightarrow
\omega +i\Gamma $. The imaginary part $\Gamma $ is the relaxation frequency
of the excitations in the homogeneous Bose gas. The value $\Gamma $ depends
on $\omega $ and for the small $\omega $ this value is much smaller than $%
\omega $, i.e., $\omega >>\Gamma $ [20]. Thus, the equation for the first
component of the field $\eta $ has the frequency $\omega $ in the form of $%
\omega \rightarrow \omega ^{\ast }=\omega -i\Gamma $ and the equation for
the second one has the frequency in the form $\omega \rightarrow \omega
=\omega +i\Gamma $.

The Keldysh Green function of this equation is denoted as $\widehat{D}\left(
x,0\right) $. We use the ''triangular'' representation. The solution of Eq.
(7.2) with the fixed boundary conditions is

\begin{equation}
\eta \left( x\right) =\widehat{D}\left( x,0\right) \left( \widehat{D}\left(
0,0\right) \right) ^{-1}\eta \left( 0\right)  \eqnum{7.3}
\end{equation}
the coordinates x=$\pm d$ of the boundary we denote as 0. The Green function 
$\widehat{D}\left( x,0\right) $ in the ''triangular'' representation reads

\begin{equation}
\widehat{D}=\left( 
\begin{array}{cc}
0 & D_{A} \\ 
D_{R} & D_{K}
\end{array}
\right)  \eqnum{7.4}
\end{equation}
where $D_{R}$ and $D_{A}$ are the retarded and advanced Green functions
respectively, and $D_{K}$ is the kinetic Green function. For the equilibrium
state, the Green function $D_{K}$ is connected with the functions $D_{R}$
and $D_{A}$ by the equation

\begin{equation}
D_{K}=\coth \left( \frac{\omega }{2T}\right) \left( D_{R}-D_{A}\right) 
\eqnum{7.5}
\end{equation}
The Green functions $D_{R}$ and $D_{A}$ have the form

\begin{eqnarray}
D_{R}\left( x,0\right) &=&\int \frac{dp_{x}}{2\pi }\frac{e^{ixp_{x}}}{%
p_{x}^{2}-\left( \omega +i\Gamma \right) ^{2}}=  \eqnum{7.6} \\
&=&\theta \left( x\right) \frac{i}{2\left( \omega +i\Gamma \right) }%
e^{ix\left( \omega +i\Gamma \right) }-\theta \left( -x\right) \frac{i}{%
2\left( \omega +i\Gamma \right) }e^{-ix\left( \omega +i\Gamma \right) } 
\nonumber
\end{eqnarray}

\begin{eqnarray*}
D_{A}\left( x,0\right) &=&\int \frac{dp_{x}}{2\pi }\frac{e^{ixp_{x}}}{%
p_{x}^{2}-\left( \omega -i\Gamma \right) ^{2}}= \\
&=&\theta \left( -x\right) \frac{i}{2\left( \omega +i\Gamma \right) }%
e^{ix\left( \omega -i\Gamma \right) }-\theta \left( x\right) \frac{i}{%
2\left( \omega +i\Gamma \right) }e^{-ix\left( \omega +i\Gamma \right) }
\end{eqnarray*}
The inverse Green function $\widehat{D}^{-1}$ in the ''triangular'' form is

\begin{equation}
\widehat{D}^{-1}=\left( 
\begin{array}{cc}
-D_{K}D_{R}^{-1}D_{A}^{-1} & D_{R}^{-1} \\ 
D_{A}^{-1} & 0
\end{array}
\right) =\left( 
\begin{array}{cc}
\coth \left( \frac{\omega }{2T}\right) \left( D_{R}^{-1}-D_{A}^{-1}\right) & 
D_{R}^{-1} \\ 
D_{A}^{-1} & 0
\end{array}
\right)  \eqnum{7.7}
\end{equation}
The space derivatives of these Green functions give (x%
\mbox{$>$}%
0)

\begin{eqnarray}
\nabla _{x}D_{R}\left( x,0\right) &=&-\frac{1}{2}e^{ix\left( \omega +i\Gamma
\right) }  \eqnum{7.8} \\
\nabla _{x}D_{A}\left( x,0\right) &=&-\frac{1}{2}e^{-ix\left( \omega
-i\Gamma \right) }  \nonumber
\end{eqnarray}

Using Eq. (7.3), we can represent the surface terms (5.12) and (4.15) of the
action $\widetilde{S}_{surf}^{\left( \varphi \right) }$ \ as

\begin{eqnarray}
&&\left[ \eta \nabla _{x}\eta \right] \left( 0\right) =  \eqnum{7.9} \\
&=&\left( 
\begin{array}{cc}
\eta _{1}\left( x\right) & \eta _{2}\left( x\right)
\end{array}
\right) \left( 
\begin{array}{cc}
0 & 1 \\ 
1 & 0
\end{array}
\right) \left( 
\begin{array}{cc}
0 & \nabla _{x}D_{A}\left( x,0\right) \\ 
\nabla _{x}D_{R}\left( x,0\right) & \nabla _{x}D_{K}\left( x,0\right)
\end{array}
\right) \left( \widehat{D}\left( 0,0\right) \right) ^{-1}\left( 
\begin{array}{c}
\eta _{1}\left( 0\right) \\ 
\eta _{2}\left( 0\right)
\end{array}
\right)  \nonumber
\end{eqnarray}

\begin{equation}
\left( 
\begin{array}{cc}
0 & \nabla _{x}D_{A}\left( 0\right) \\ 
\nabla _{x}D_{R}\left( 0\right) & \nabla _{x}D_{K}\left( 0\right)
\end{array}
\right) =-\frac{1}{2}\left( 
\begin{array}{cc}
0 & 1 \\ 
1 & 0
\end{array}
\right)  \eqnum{7.10}
\end{equation}

\begin{eqnarray}
\left( \widehat{D}\left( 0,0\right) \right) ^{-1} &=&\left( 
\begin{array}{cc}
\coth \left( \frac{\omega }{2T}\right) \left( D_{R}^{-1}\left( 0,0\right)
-D_{A}^{-1}\left( 0,0\right) \right) & D_{R}^{-1}\left( 0,0\right) \\ 
D_{A}^{-1}\left( 0,0\right) & 0
\end{array}
\right) =  \eqnum{7.11} \\
&=&2\left( 
\begin{array}{cc}
-2i\omega \coth \left( \frac{\omega }{2T}\right) & -i\omega \\ 
i\omega & 0
\end{array}
\right)  \nonumber
\end{eqnarray}
Thus, we have

\begin{equation}
\left[ \eta \nabla _{x}\eta \right] \left( 0\right) =\left( 
\begin{array}{cc}
\eta _{1}\left( 0\right) & \eta _{2}\left( 0\right)
\end{array}
\right) \left( 
\begin{array}{cc}
-2i\omega \coth \left( \frac{\omega }{2T}\right) & -i\omega \\ 
i\omega & 0
\end{array}
\right) \left( 
\begin{array}{c}
\eta _{1}\left( 0\right) \\ 
\eta _{2}\left( 0\right)
\end{array}
\right)  \eqnum{7.12}
\end{equation}
Substituting (7.12) into (4.15),\ we obtain the following expression for the
action $\widetilde{S}_{surf}^{\left( \varphi \right) }$ in the ''triangle''
representation

\begin{equation}
\widetilde{S}_{surf}^{\left( \varphi \right) }=-\frac{1}{8}n_{0}\xi
^{3}S_{\perp }\int dt\left\{ \left( 
\begin{array}{cc}
\phi _{s}, & \Phi _{s}
\end{array}
\right) \left( \widehat{D}\left( 0,0\right) \right) ^{-1}\left( 
\begin{array}{c}
\phi _{s} \\ 
\Phi _{s}
\end{array}
\right) \right\}  \eqnum{7.13}
\end{equation}
In the time representation the frequency $\omega $ is replaced by the time
derivative $i\partial _{t}$. The fields $\Phi $ and $\phi $ are defined by
the equalities

\begin{eqnarray}
\Phi &=&\frac{1}{2}\left( \varphi _{1}+\varphi _{2}\right)  \eqnum{7.14} \\
\phi &=&\frac{1}{2}\left( \varphi _{1}-\varphi _{2}\right)  \nonumber
\end{eqnarray}
where $\varphi _{1}$ and $\varphi _{2}$ are the phase fields determined on
the upper and the lower brunches of the time contour, respectively.

As a result, the action $S_{surf}^{\left( \varphi \right) }$ \ as an
effective action $S_{eff}$ which describes the relaxation of the difference
between the phases at the boundary of the region of the nonzero external
potential can be written in the ''triangular'' form as

\begin{eqnarray}
S_{eff} &=&S_{surf}^{\left( \varphi \right) }=  \eqnum{7.15} \\
&=&n_{0}\xi ^{3}S_{\perp }\int dt\left\{ 
\begin{array}{c}
\sum\limits_{\alpha =1,2}\left( 
\begin{array}{cc}
\phi _{s\alpha }, & \Phi _{s\alpha }
\end{array}
\right) \left[ \left( \widehat{G}_{s}^{\left( 0\right) }\right) ^{-1}-\frac{1%
}{8}\left( \widehat{D}\left( 0,0\right) \right) ^{-1}\right] \left( 
\begin{array}{c}
\phi _{s\alpha } \\ 
\Phi _{s\alpha }
\end{array}
\right) - \\ 
-\left( V\left[ \Delta \Phi _{s}+\Delta \phi _{s}\right] -V\left[ \Delta
\Phi _{s}-\Delta \phi _{s}\right] \right)
\end{array}
\right\}  \nonumber
\end{eqnarray}
where

\begin{equation}
\left( \widehat{G}_{s}^{\left( 0\right) }\right) ^{-1}=\frac{1}{2\varkappa }%
\left( 
\begin{array}{cc}
0 & \omega ^{2} \\ 
\omega ^{2} & 0
\end{array}
\right)  \eqnum{7.16}
\end{equation}

\begin{equation}
V\left( \varphi _{s}\right) =-\frac{1}{\varkappa \sinh \left( 2\varkappa
d\right) }\cos \left( \varphi _{s}\right)  \eqnum{7.17}
\end{equation}
The index $\alpha =1,2$ in Eq. (7.15) corresponds to the right-hand or the
left-hand side of the bulk and $\Delta \Phi _{s}=\Phi _{s1}-\Phi _{s2}$, $%
\Delta \phi _{s}=\phi _{s1}-\phi _{s2}$. The value $S_{\perp }$ is the area
of the section of the bulk along the y-z plane.

\section{The description of the relaxation}

The effective action (7.15) describes the relaxation of the difference
between the phases at the right and the left boundaries of the region of the
nonzero external potential. The generation functional Z can be written in
the form

\begin{equation}
Z=\int D\phi D\Phi e^{i\left( S_{eff}+S_{j}\right) }  \eqnum{8.1}
\end{equation}
To simplify the consideration, we suppose that the temperature T is much
larger than the characteristic frequency $\omega _{0}\thicksim \gamma
=\left( \varkappa \sinh \left( 2\varkappa d\right) \right) ^{-1}$. In this
assumption, firstly, the term $\coth \left( \frac{\omega }{2T}\right) $ in $%
\left( \widehat{D}\left( 0,0\right) \right) ^{-1}$ (7.11) can be written as $%
\frac{2T}{\omega }$. Secondly, the component of the phase field $\phi $
describing the quantum fluctuations is much smaller than the component $\Phi 
$ ($\phi >>\Phi $). This approximation is conditioned by the large value of
the parameter $n_{0}\xi ^{3}S_{\perp }>>1$ (the system of units is such that
the scale of the length is defined by the value $\xi $). Due to the
smallness of the field $\phi $ compared with the field $\Phi $ it is
possible to expand the nonlinear terms $V\left( \Delta \Phi _{s}\pm \Delta
\phi _{s}\right) $ in a series in $\Delta \phi _{s}$ to first order. Thus,
we obtain

\begin{equation}
S_{eff}=n_{0}\xi ^{3}S_{\perp }\int dt\left\{ 
\begin{array}{c}
\sum\limits_{\alpha =1,2}\phi _{s\alpha }\left[ \frac{1}{2\varkappa }\omega
^{2}+\frac{i}{4}\omega \right] \Phi _{s\alpha }+\Phi _{s\alpha }\left[ \frac{%
1}{2\varkappa }\omega ^{2}-\frac{i}{4}\omega \right] \phi _{s\alpha }+i\phi
_{s\alpha }T\phi _{s\alpha }- \\ 
-2\left( \Delta \phi _{s}\right) V^{\prime }\left[ \Delta \Phi _{s}\right]
\end{array}
\right\}  \eqnum{8.2}
\end{equation}
The term $V^{\prime }\left( \Delta \Phi _{s}\right) $ is the functional
derivative of the functional $V\left[ \Delta \Phi _{s}\right] $, i.e., $%
V^{\prime }\left[ \Delta \Phi _{s}\right] =\delta V\left[ \Delta \Phi _{s}%
\right] /\delta \left( \Delta \Phi _{s}\right) $.

We introduce the half-sum and the half-difference of the fields $\phi
_{s\alpha }$ and $\Phi _{s\alpha }$

\begin{eqnarray}
\widetilde{\Phi } &=&\frac{1}{2}\left( \Phi _{s1}+\Phi _{s2}\right) \text{
;\ \ \ \ \ \ \ \ \ \ \ \ }\widetilde{\phi }=\frac{1}{2}\left( \phi
_{s1}+\phi _{s2}\right)  \eqnum{8.3} \\
\left( \Delta \Phi \right) &=&\frac{1}{2}\left( \Phi _{s1}-\Phi _{s2}\right) 
\text{ ;\ \ \ \ \ \ \ \ \ \ \ \ }\left( \Delta \phi \right) =\frac{1}{2}%
\left( \phi _{s1}-\phi _{s2}\right)  \nonumber
\end{eqnarray}
After the integration of the generation potential Z over the fields $%
\widetilde{\Phi }$ and $\widetilde{\phi }$ we obtain

\begin{equation}
Z=\int D\left( \Delta \phi \right) D\left( \Delta \Phi \right) e^{i\left(
\Delta S_{eff}+S_{j}\right) }  \eqnum{8.4}
\end{equation}
where

\begin{equation}
\Delta S_{eff}=n_{0}\xi ^{3}S_{\perp }\int dt\left\{ 
\begin{array}{c}
\left( \Delta \phi \right) \left[ \frac{1}{2\varkappa }\omega ^{2}+\frac{i}{4%
}\omega \right] \left( \Delta \Phi \right) +\left( \Delta \Phi \right) \left[
\frac{1}{2\varkappa }\omega ^{2}-\frac{i}{4}\omega \right] \left( \Delta
\phi \right) + \\ 
+i\left( \Delta \phi \right) T\left( \Delta \phi \right) -2\left( \Delta
\phi \right) V^{\prime }\left[ \Delta \Phi \right]
\end{array}
\right\}  \eqnum{8.5}
\end{equation}

The fields $\Delta \Phi $ and $\Delta \phi $ are represented as a sum of the
parts changing slowly and fast in time. We denote the fast parts of $\Delta
\Phi $ and $\Delta \phi $ as $\delta \Phi $ and $\delta \phi $ and the slow
parts as $\Phi $ and $\phi $. Thus

\begin{eqnarray}
\Delta \Phi &\rightarrow &\Phi +\delta \Phi  \eqnum{8.6} \\
\Delta \phi &\rightarrow &\phi +\delta \phi  \nonumber
\end{eqnarray}
The criterion of separating the fields into the fast and slow ones is the
magnitude of the frequency. If the frequency of the field $\omega \gtrsim
\omega _{c}$ where $\omega _{c}=1/\varkappa $, the field is fast,
correspondingly, if the frequency $\omega <<\omega _{c}$, the field is slow.
This gives

\begin{equation}
S_{eff}=S_{eff}^{\left( slow\right) }+S_{eff}^{\left( fast\right) } 
\eqnum{8.7}
\end{equation}

\begin{equation}
S_{eff}^{\left( slow\right) }=n_{0}\xi ^{3}S_{\perp }\int dt\left\{ 
\begin{array}{c}
\phi \left[ \frac{1}{2\varkappa }\omega ^{2}+\frac{i}{4}\omega \right] \Phi
+\Phi \left[ \frac{1}{2\varkappa }\omega ^{2}-\frac{i}{4}\omega \right] \phi
+ \\ 
+i\phi T\phi -2\phi V^{\prime }\left[ \Phi \right]
\end{array}
\right\}  \eqnum{8.8}
\end{equation}

\begin{equation}
S_{eff}^{\left( fast\right) }=n_{0}\xi ^{3}S_{\perp }\int dt\left\{ 
\begin{array}{c}
\left( \delta \phi \right) \left[ \frac{1}{2\varkappa }\omega ^{2}+\frac{i}{4%
}\omega -V^{\prime \prime }\left[ \Phi \right] \right] \left( \delta \Phi
\right) + \\ 
\left( \delta \Phi \right) \left[ \frac{1}{2\varkappa }\omega ^{2}-\frac{i}{4%
}\omega -V^{\prime \prime }\left[ \Phi \right] \right] \left( \delta \phi
\right) +i\left( \delta \phi \right) T\left( \delta \phi \right)
\end{array}
\right\}  \eqnum{8.9}
\end{equation}
The integration over $\delta \phi $ yields

\begin{equation}
S_{\Phi }^{\left( fast\right) }=n_{0}\xi ^{3}S_{\perp }\int dt\left\{ i\frac{%
1}{16T}\left[ \widehat{L}\left( \delta \Phi \right) \right] ^{2}\right\} 
\eqnum{8.10}
\end{equation}
where the operator $\widehat{L}$ is

\begin{equation}
\widehat{L}=\frac{1}{2\varkappa }\partial _{t}^{2}+\frac{1}{4}\partial
_{t}+V^{\prime \prime }\left( \Phi \right)  \eqnum{8.11}
\end{equation}
Note that for the slow frequencies $\omega <<1$, the term $\frac{1}{%
2\varkappa }\partial _{t}^{2}$ can be neglected compared with the term $%
\frac{1}{4}\partial _{t}$.

For the generation potential, we obtain

\begin{equation}
Z=Z_{slow}Z_{fast}  \eqnum{8.12}
\end{equation}
where

\begin{equation}
Z_{slow}=\int D\phi D\Phi e^{i\left( S_{eff}^{\left( slow\right)
}+S_{j}^{\left( slow\right) }\right) }  \eqnum{8.13}
\end{equation}

\begin{equation}
Z_{fast}=\int D\left( \delta \Phi \right) e^{i\left( S_{\Phi }^{\left(
fast\right) }+S_{j}^{\left( fast\right) }\right) }  \eqnum{8.14}
\end{equation}
Due to the form of the action $S_{\Phi }^{\left( fast\right) }$ (8.10) the
generation potential $Z_{fast}$ has the Gaussian form and can be calculated

\begin{equation}
Z_{fast}=C_{1}det_{1}\left[ \widehat{L}^{-1}\right]  \eqnum{8.15}
\end{equation}
where the constant C has the form

\[
C_{1}=\int D\left( \delta \Phi \right) \exp \left\{ -\int dt\left\{ \frac{1}{%
16T}\left( \delta \Phi \right) ^{2}\right\} \right\} 
\]
and can be included into a measure of the functional integral $Z_{slow}$.
The index 1 in det$_{1}$ denotes the product of the Fourier components
corresponding to the fast frequencies. The value $det_{1}\left[ \widehat{L}%
^{-1}\right] $ can be calculated by the following way

\begin{eqnarray}
det_{1}\left[ \widehat{L}^{-1}\right] &=&e^{Sp^{\prime \prime }Ln\left[ 
\widehat{L}^{-1}\right] }=C_{2}\exp \left\{ Sp^{\prime \prime }Ln\left[
\left( G_{R}^{\left( 0\right) }\right) V^{\prime \prime }\left( \Phi \right) %
\right] \right\} =  \eqnum{8.16} \\
&=&C_{2}\exp \left\{ -Sp^{\prime }Ln\left[ \left( G_{R}^{\left( 0\right)
}\right) V^{\prime \prime }\left( \Phi \right) \right] \right\}  \nonumber
\end{eqnarray}
The symbol $Sp^{\prime \prime }$ indicates the sum over the fast frequencies
and the symbol $Sp^{\prime }$ indicates the sum over the slow frequencies.
The inessential c-number constant $C_{2}$ is equal to

\[
C_{2}=\exp \left\{ -Sp^{\prime \prime }Ln\left[ \left( G_{R}^{\left(
0\right) }\right) ^{-1}\right] \right\} 
\]
and can be included into the measure of the $Z_{slow}$. The value $%
Sp^{\prime }Ln\left[ \left( G_{R}^{\left( 0\right) }\right) V^{\prime \prime
}\left( \Phi \right) \right] $ is calculated as

\[
Sp^{\prime }Ln\left[ \left( G_{R}^{\left( 0\right) }\right) V^{\prime \prime
}\left( \Phi \right) \right] =V^{\prime \prime }\left( \Phi \left( t\right)
\right) \int\limits_{-\frac{1}{4}}^{\frac{1}{4}}\frac{d\omega }{2\pi }\frac{1%
}{\frac{i}{4}\left( \omega +i\Gamma \right) }=-4\theta \left( 0\right)
V^{\prime \prime }\left( \Phi \right) =-2V^{\prime \prime }\left( \Phi
\right) 
\]
Hence, the expression for the generation potential has the form

\begin{equation}
Z_{fast}=C_{1}det_{1}\left[ \widehat{L}^{-1}\right] =\exp \left\{ 2\int
dtV^{\prime \prime }\left( \Phi \right) \right\}  \eqnum{8.17}
\end{equation}
This equation gives

\begin{equation}
Z=Z_{slow}Z_{fast}=\int D\phi D\Phi e^{i\left( S_{eff}^{\left( slow\right)
}+S_{j}^{\left( slow\right) }-2i\int dtV^{\prime \prime }\left( \Phi \right)
\right) }  \eqnum{8.18}
\end{equation}
The action $S_{eff}^{\left( fast\right) }$ has form (8.9). Integrating Z
(8.18) over the field $\phi $, we obtain

\begin{equation}
Z=\int D\Phi e^{-\left( S_{slow}^{\left( \Phi \right) }-2\int dtV^{\prime
\prime }\left( \Phi \right) -iS_{j}^{\left( slow\right) }\right) } 
\eqnum{8.19}
\end{equation}

\begin{equation}
S_{slow}^{\left( \Phi \right) }=\int dt\left\{ \frac{1}{T}\left[ \frac{1}{4}%
\left( \partial _{t}\Phi \right) -V^{\prime }\left( \Phi \right) \right]
^{2}-2V^{\prime \prime }\left( \Phi \right) \right\}  \eqnum{8.20}
\end{equation}
Omitting the term $\left( \partial _{t}\Phi \right) V^{\prime }\left( \Phi
\right) $ which can be represented as a total derivative, we obtain

\begin{equation}
S_{slow}^{\left( \Phi \right) }=\frac{1}{T}\int dt\left\{ \frac{1}{16}\left(
\partial _{t}\Phi \right) ^{2}+\left( V^{\prime }\left( \Phi \right) \right)
^{2}-2TV^{\prime \prime }\left( \Phi \right) \right\}  \eqnum{8.21}
\end{equation}

Note that the generation potential Z with action $S_{slow}^{\left( \Phi
\right) }$ (8.21) has the form corresponding to the quantum mechanics with
the ''imaginary time'' in which the role of the ''imaginary time'' plays the
usual real time. This action corresponds to the action which is obtained for
the classical random process describing by the Langeven equation with the
white noise [21] and having the form of the supersymmentic quantum
mechanics. 

The Hamiltonian corresponding to the action $S_{slow}^{\left( \Phi \right) }$
(8.21) is

\begin{equation}
\widehat{H}=-4T^{2}\frac{d^{2}}{d\Phi ^{2}}+\left( V^{\prime }\left( \Phi
\right) \right) ^{2}-2TV^{\prime \prime }\left( \Phi \right)  \eqnum{8.22}
\end{equation}
and the Schredinger equation corresponding to this Hamiltonian can be
written as

\begin{equation}
-\partial _{t}\Psi =\widehat{H}\Psi  \eqnum{8.23}
\end{equation}
The Hamiltonian $\widehat{H}$ can be represented in the factorized form

\begin{equation}
\widehat{H}=-\left( 2T\right) ^{2}\left( \frac{d}{d\Phi }-A\right) \left( 
\frac{d}{d\Phi }+A\right)  \eqnum{8.24}
\end{equation}
where

\begin{equation}
A=\frac{V^{\prime }\left( \Phi \right) }{2T}  \eqnum{8.25}
\end{equation}

The Hamiltonian $\widehat{H}$ has the minimum eigenvalue equal to zero. The
eigenwave function corresponding to the zero eigenvalue is

\begin{equation}
\Psi _{0}=\exp \left\{ \int d\Phi A\left( \Phi \right) \right\} =\exp
\left\{ -\frac{V\left( \Phi \right) }{2T}\right\}  \eqnum{8.26}
\end{equation}

This can easily be seen from the form of the Hamiltonian (8.24). Due to the
form of the Hamiltonian (8.24) the Schredinger Eq. (8.23) can be modified
into the form of the Fokker-Planck equation [18] by the transformation

\begin{equation}
\Psi \rightarrow \exp \left\{ -\int\limits_{0}^{\Phi }d\Phi ^{\prime
}A\left( \Phi ^{\prime }\right) \right\} \Psi  \eqnum{8.27}
\end{equation}
In this case the eigenwave function (8.26) corresponds to the equilibrium
distribution function for the field $\Phi $ which is formed as a result of
the relaxation process.

This research was supported by the Russian Foundation for Basic Research and
by the Netherlands Organization for Scientific Research (NWO).

\end{document}